\documentclass{article}

 \usepackage[preprint]{neurips_2020}

\usepackage[utf8]{inputenc} % allow utf-8 input
\usepackage[T1]{fontenc}    % use 8-bit T1 fonts
\usepackage{hyperref}       % hyperlinks
\usepackage{url}            % simple URL typesetting
\usepackage{booktabs}       % professional-quality tables
\usepackage{amsfonts}       % blackboard math symbols
\usepackage{nicefrac}       % compact symbols for 1/2, etc.
\usepackage{microtype}      % microtypography
\usepackage{graphicx}
\usepackage{amsmath}
\usepackage{amssymb}
\usepackage{subcaption}
\usepackage{natbib}
\usepackage{verbatim}
\usepackage{wrapfig}

\usepackage{amsthm,dsfont}
\usepackage{gensymb}
\usepackage{bbm}
\usepackage{wasysym}
\usepackage{float}
\usepackage[normalem]{ulem}

\usepackage{xcolor}

\title{Multiview Variational Deep Learning with Application to Scalable Indoor Localization}

\author{
  Minseuk Kim \\
  School of Electrical Engineering\\
  KAIST, Daejeon, Republic of Korea\\
  \texttt{kms4105@kaist.ac.kr} \\
  \And
  Changjun Kim \\
  School of Electrical Engineering\\
  KAIST, Daejeon, Republic of Korea\\
  \texttt{changjun0605@kaist.ac.kr} \\
  \AND
  Dongsoo Han\\
  School of Computing\\
  KAIST, Daejeon, Republic of Korea\\
  \texttt{dshan@kaist.ac.kr} \\
  \And
  June-Koo Kevin Rhee\\
  School of Electrical Engineering\\
  ITRC of Quantum Computing for AI\\
  KAIST, Daejeon, Republic of Korea\\
  \texttt{rhee.jk@kaist.edu}\\
}

\begin{document}

\maketitle

\begin{abstract}
Radio channel state information (CSI) measured with many receivers is a good resource for localizing a transmit device with machine learning with a discriminative model. However, CSI localization is nontrivial when the radio map is complicated, such as in building corridors. This paper introduces a view-selective deep learning (VSDL) system for indoor localization using CSI of WiFi. The multiview training with CSI obtained from multiple groups of access points (APs) generates latent features on supervised variational deep network. This information is then applied to an additional network for dominant view classification to enhance the regression accuracy of localization. As non-informative latent features from multiple views are rejected, we can achieve a localization accuracy of 0.77 m, which outperforms by 30 $\%$ the best known accuracy in practical applications in a real building environment. To the best of our knowledge, this is the first approach to apply variational inference and to construct a scalable system for radio localization. Furthermore, our work investigates a methodology for supervised learning with multiview data where informative and non-informative views coexist. 

%Many machine learing algorithms have been developed for accurate localization system in wireless indoor area. Channel State Information (CSI),
%which has become a measurement alternative rather than Received Singal Strength Indicator (RSSI), is a multi-dimension data according to different channel frequencies. In this paper, we propose a novel variational deep learning system with application to indoor localization using CSI. Extracted latent features from variational inferences properly represents the CSI data characteristics. As well as advanced localization accuracy, we can localize the target even at the multi-corridor environment through our prior knolwedge-based supervised corridor selective learning method and corridor classification fusion network that optimizes both corridor classification and desired regression. 
%We demonstrate the better performance compared to existing machine learning methods, which is under $1m$ localization accuracy on a two-corridor environment. This is the first approach to apply variational inference and to construct scalable system on dynamic environment in the indoor localization area.  

\end{abstract}

\section{Introduction}

%To discriminate a desired output from the multiview data, the strategy can be divided into three categories according to how early the system combines the output from each view \cite{noble2004support}: 1) early integration which combines all views first and extracts the correlated feature at once, 2) late integration, which extracts features from each view and concatenates at the decision level, and 3) intermediate integration, which concatenates the intermediate output at the learning level. Depending on how the multiview data are correlated, one can select the appropriate integration.

Machine learning applications with a multiview embodiment rendered with multiple sources can exploit feature correlations among various views to attain the best model for inference. Multiview data can be efficiently generated by variational inference \cite{hoffman2013stochastic} from a single view, as reported in \cite{zhao2018multi}. Variational inference is also adopted in broad discriminative models such as clustering \cite{dilokthanakul2016deep}, classification \cite{xu2017variational}, and regression \cite{girolami2006variational} to utilize the probabilistic latent feature space. From the perspective of a deep network, variational deep learning (DL) \cite{wilson2016stochastic} can jointly optimize objectives of a Gaussian process marginal likelihood to train the deep network. Reparameterization of variational inference \cite{wainwright2008graphical} derives mean-field latent feature vectors to represent the posterior of an input. Compared to stand-alone deep networks, support vector machines (SVMs), and other recent systems, the variational DL can substantially improve classification and regression performance \cite{wainwright2008graphical}. In this paper, we apply the variational DL to localization of transmit devices based on radio signal data in a practical WiFi environment. 

Similarly to an example that generates a single image view from multiple image views \cite{tulsiani2017multi}, radio signals retrieved in multiple groups of receivers at distributed locations can form multiview information for machine learning to locate the transmit device. By the nature of variational DL that the elements of the extracted latent vector have no correlations, we can map between a multivariate standard normal distribution and each view. As the variational DL is forced to ensure its latent vector approximated to the standard normal distribution with a reduced number of independent Gaussian processes, we can mitigate the signal uncertainties. Compared to the global positioning system (GPS), which leads to 5 m to 10 m outdoor localization accuracy, indoor localization requires more accurate positioning of the transmit devices. Recently, a channel state information (CSI) of WiFi has emerged as the strong candidate for indoor localization rather than a scalar-valued received signal strength indicator (RSSI) \cite{li2018indoor}. The subcarrier CSIs of WiFi orthogonal frequency division multiplexing (OFDM) channel form complex vector information, providing much rich information on radio localization with an excellent localization accuracy.

\subsection{Related work}
At a server side of a localization system, CSI data is collected from multiple receivers at the same time to find the transmit device location. Transmit device localization is achieved by geometric analysis of such CSI data to determine the time of flight and angle of arrival information of the radio packet \cite{xiong2013arraytrack, kotaru2015spotfi}. However, in a practical indoor environment, the noise and signal fading problems become critical against finding true transmit device locations with such analytical methods. 

In order to cope with noise and signal fading, many machine learning methods have been developed to successfully find the transmit device location from the complex CSI data by considering it as a single view. References \cite{wang2015deepfi,li2019defe,wang2017biloc} utilized restricted Boltzmann machine (RBM) based approaches to reconstruct the CSI data for better likelihood determination for localization. Convolutional neural network (CNN) based approaches were proposed in \cite{berruet2018delfin,chen2017confi,wang2017cifi}. The consecutive data packets were concatenated to a single batch as a 2-D CSI image. However, in the radio localization, it is preferable that the system be capable of packet-by-packet processing, rather than batch processing. References \cite{chen2019smartphone} and \cite{zhou2017csi} introduced SVM based classification and regression, respectively. Reference \cite{gao2019crisloc} adopted transfer learning to reconstruct CSI data and applied the enhanced $k$-nearest neighbor (KNN) approach for localization. To enhance the accuracy of spot location classification, \cite{tsai2018refined} introduced autoencoder. In \cite{dang2019novel}, principal component analysis (PCA), one of the preprocessing methods, was applied to reduce the multi dimension CSI before passing through a deep neural network (DNN). Also, combining RSSI and CSI, \cite{hsieh2019deep} proposed multi-layer perceptron (MLP) and 1-D CNN. From the perspective of device-free indoor localization, \cite{sanam2020multi} carried out a canonical correlation analysis (CCA) to classify the location of a device with a human, where the device is neither a transmitter nor a receiver. The localization accuracies achieved by all the papers listed above did not reach below a meter in practical application environments, except for the case where the training and test locations are the same. However, our result in this paper has achieved a localization accuracy of sub-meter by adopting multiview architecture with variational DL. 

\subsection{Our contributions}
An advanced scalable localization with a keen accuracy is pursued in this work, to extend the area of radio localization on a complicated floor plan for an in-building application. To construct a scalable learning system for localization in a real building environment with corridors, we propose a supervised learning system named view-selective deep learning (VSDL) with CSI data consisting of multiple views. The VSDL obtains much improved regression performance due to latent feature generation by the use of the variational inference and non-informative latent feature rejection among the multiple views. The proposed VSDL achieved a localization accuracy of 0.77 m, which outperforms by more than 30\% the best known accuracy of other works applied in practical building experiment. To the best of our knowledge, this is the first approach to apply variational inference for CSI-based WiFi localization and to construct a scalable system for localization in a wide and complex environment.
Furthermore, application of our system can be extended to general supervised learning with multiview data where informative and non-informative views coexist.

%Our aim is to obtain more advanced localization accuracy using variational inference. Beyond the grid network environment, we propose a novel supervised learning system named view-selective deep learning (VSDL) in a corridor-existing real building environment with multiview CSI data. In such a scenario, we first propose a view-oriented variational deep network utilizing given information whether or not each view is informative. The sampled latent vectors from the view-oriented network are intermediately integrated and then applied to a view-classified regression network that jointly optimizes view classification and desired regression objectives. The result of view classification becomes a reweight parameter that selects the dominant view to consider, and hence further improves the regression performance. As non-informative latent features from multiple views are rejected, we can achieve a localization accuracy of 0.77 m, which surpasses the best known accuracy in practical applications in a real building environment. To the best of our knowledge, this is the first approach to apply variational inference and to construct a scalable system for localization in a complex environment. Furthermore, our system can be applied to general supervised learning with multiview data where informative and non-informative views coexist. 

\section{CSI Preliminaries and data collection}
\label{sec:CSI Preliminiaries and data collection}

In a WiFi network, device localization can be achieved by analyzing the CSIs of a radio packet arriving at multiple receiver antennas of an access point (AP), complying with the IEEE 802.11a/g/n/ac standards for the multi-input multi-output (MIMO) air interface. In the experiment with an Intel WiFi link (IWL) 5300 network interface controller (NIC), the physical layer API reports a complex CSI vector of 30 selected subcarriers for an antenna receiving a WiFi packet \cite{halperin2011predictable}. The phase difference of CSIs of multiple antennas provides angle of arrival of a received packet, which is the key information to localize the transmit device.

The received CSI of packet at subcarrier $i\in \{1, \dots ,I\}$ of antenna $m\in \{1,\dots,M\}$ with nominal CSI $H_{m,i}$ and noise $N_{m,i}$ is represented as $\hat{H}_{m,i} = |H_{m,i}|e^{j2\pi\angle{H_{m,i}}}+N_{m,i}$. The nominal CSI amplitude is $|H_{m,i}|$, and the nominal CSI phase is represented as
\begin{equation}\label{eq:1}
 \angle{H_{m,i}} = s_i  ~\delta_f~\tau +(m-1){f_c}\frac{d\sin\theta}{c},
\end{equation}
where a subset of subcarriers {$\{s_i\}$} is selected among available subcarriers indexed between $-S$ and $+S$. The constant $\delta_f$ is the frequency difference between subcarriers, $f_c$ is the center frequency of the channel, $d$ is the distance between adjacent receiver antennas, and $c$ is the speed of light. Here, the phase $\angle{H_{m,i}}$ is a function of the time of flight $\tau$ and the angle of arrival $\theta$, which implies that subcarrier frequencies and the geometry of the antenna array cause the relative phase difference due to different radio arrival times. Many of the previous localization techniques aimed to find the nominal time of flight and angle of arrival \cite{schmidt1986multiple}. But in real 802.11 communication, several offsets and noise accompany them, and the measured phase $\angle{\hat{H}}_{m,i}$ is represented as
%\begin{equation}\label{eq:2}
$\angle{\hat{H}}_{m,i} = \angle{H_{m,i}}  + s_i~\lambda+\mu_m+\beta+Z_{m,i},$
%\end{equation}
where $\lambda$ and $\mu_m$ denote the subcarrier-dependent offset coefficient and the receiver antenna-dependent offset, respectively, and $\beta$ and $Z_{m,i}$ denote packet-dependent offset and noise, respectively \cite{tzur2015direction}. Empirically, these offset and noise cause a large fluctuation to the CSI phase and thus make it hard to be solved by the analytical methods. In our approach with variational inference, the CSI is mapped to an unit phasor complex that measures phase difference of CSIs among different antennas:
\begin{equation}
    x_{m,i}=\frac{\hat{H}_{m,i}/|\hat{H}_{m,i}|}{\hat{H}_{M,i}/|\hat{H}_{M,i}|}, ~~~~m\in\{1,\dots,M-1\}.
\end{equation}
Here, we adopt variational DL to mitigate the noise and signal fading problems in the CSI training samples, defined as $\mathbf{x}=[x_{m,i}]$, with $m\in\{1,\dots,M-1\}$ and $i\in\{1,\dots,I\}$.

In order to construct the localization system scalable in a complex area, we should consider exclusion of non-informative CSI views. One can deploy multiple APs over the area consisting of $K$ sub-areas, where a collection of APs in each sub-area forms a view of the CSI sample. We apply deep learning to classify the dominant views from multiple sub-areas. 

\begin{comment}
Here we can take the phase difference between adjacent antennas to remove $\tau$, $\lambda$, and $\epsilon$ as
\begin{equation}\label{eq:3}
\begin{split}
 \triangle\angle{\hat{H}}_{m,i}
& = \angle\frac{{\hat{H}}_{m+1,i}}{{\hat{H}}_{m,i}}\\
& =  \angle{H_{m,i}}+ \mu_{m+1}-\mu_m+\triangle\beta+\triangle{Z}\\
&= {f_c}\frac{d\sin\theta}{c}+ \mu_{m+1}-\mu_m+\triangle\beta+\triangle{Z}.\\
\end{split}
\end{equation} 
For each antenna pair $(m,m+1)$, $\triangle\angle{\hat{H}}_{m,i}$ must be consistent without noise for every subcarrier $i$. Therefore we use the phase difference information as learning input. By considering multiple access points (APs) as receivers which are equipped with multiple antennas, the CSI phase difference information has single dimension but is a combination of multiple probability distributions. With $V$ receivers, the multiview learning input is a collection of all information, $\mathbf{x}^{\mathbf{I}\times\mathbf{(M-1)}\times\mathbf{V}}$, for every sample.
Now we will introduce the variational DL to properly exploits its multiview characteristics and finds the latent vectors in the following section. 
\end{comment}
 
\section{Variational Deep Learning}
\label{sec:Variational Deep Learning}

We apply a variational DL for regression to be trained with input as a pair of $\mathbf{x}$ (i.e., CSI in our case) and a true label $\mathbf{y}$ (i.e., Cartesian coordinate in our case). Let us assume a latent feature vector $\mathbf{z}$ consisting of $z_j=f_j(\mathbf{x}), j\in\{1,\dots,J\}$ of independent Gaussian processes (GPs) with probability of $z_j \sim \mathcal{GP}(\mu_j,\sigma_j^2)$, where the mean-field feature vectors $\boldsymbol{\mu} = [\mu_1,\dots,\mu_J]$ and $\boldsymbol{\sigma} = [\sigma_1,\dots,\sigma_J]$ indicate mean and standard deviation, respectively. The latent vector is used to estimate the regression output $\mathbf{\hat{y}}$ where the true $\mathbf{y}$ is supposed to be represented as $\mathbf{y}(\mathbf{x})|\mathbf{z} \sim \mathcal{N}(\mathbf{y}(\mathbf{x});\mathbf{z},\boldsymbol{\sigma}^2\odot\boldsymbol{I})$. In order to make the learning variables differentiable for gradient descent based back-propagation, it requires reparameterization. The weights and biases of a neural network (NN) to obtain the latent vector $\mathbf{z}$ from the input $\mathbf{x}$ is updated by sampling of noise vector $\boldsymbol{\epsilon}$:
\begin{equation}\label{eq:4}
\mathbf{z} = \boldsymbol{\mu}+\boldsymbol{\sigma}\odot\boldsymbol{\epsilon}, ~~~\boldsymbol{\epsilon} \sim \mathcal{N}(0,\mathbf{I}).    
\end{equation}
Then with the variational posterior over the estimated distribution $q(\mathbf{z})$, Jensen's inequality can be applied to give evidence lowerbound (ELBO) of the marginal log-likelihood of regression \cite{wilson2016stochastic}:
\begin{equation}\label{eq:5}
 \log p(\mathbf{y}) \geq \mathbb{E}_{q(\mathbf{z}|\mathbf{y})}[\log p(\mathbf{y}|\mathbf{z})]
-\mathbf{KL}[q(\mathbf{z}|\mathbf{y})||p(\mathbf{z})] \triangleq \mathcal{L}(q) = \mathbf{ELBO}.
\end{equation}
In (\ref{eq:5}), the first term at the right hand side is cross entropy between the label $\mathbf{y}$ and the regression output from the latent vector $\mathbf{z}$, which is equivalent to the regression loss. The second term is the Kullback–Leibler (KL) divergence between the estimated posterior $q(\mathbf{z}|\mathbf{y})$ and the distribution $p(\mathbf{z})$. Our aim is to approximate the estimated posterior $q(\mathbf{z}|\mathbf{y})$ to be close to the posterior $p(\mathbf{z}|\mathbf{y})$; in other words, to minimize $\mathbf{KL}[q(\mathbf{z}|\mathbf{y})||p(\mathbf{z}|\mathbf{y})]$. Again the log-likelihood can be represented as the following equation:
\begin{equation}\label{eq:6}
\begin{split}
\log p(\mathbf{y})  
& = \int \log \big (p(\mathbf{y}) \big) q(\mathbf{z}|\mathbf{y})dz \\
& = \int \log \frac{p(\mathbf{y},\mathbf{z})}{q(\mathbf{z}|\mathbf{y})}q(\mathbf{z}|\mathbf{y})dz 
+ \int \log \frac{q(\mathbf{z}|\mathbf{y})}{p(\mathbf{z}|\mathbf{y})}q(\mathbf{z}|\mathbf{y})dz \\
& = \mathbf{ELBO} + \mathbf{KL}[q(\mathbf{z}|\mathbf{y})||p(\mathbf{z}|\mathbf{y})].
\end{split}
\end{equation} 
Since the log-likelihood $\log p(\mathbf{y})$ is bounded, in order to minimize the KL divergence $\mathbf{KL}[q(\mathbf{z}|\mathbf{y})||p(\mathbf{z}|\mathbf{y})]$, we have to maximize the term ELBO. The KL divergence term in (\ref{eq:5}) with the normal distribution $q(\mathbf{z}|\mathbf{y})$ and the standard normal distribution $p(\mathbf{z})$ can be simplified to 
\begin{equation}\label{eq:7}
\mathbf{KL}[q(\mathbf{z}|\mathbf{y})||p(\mathbf{z})] = \frac{1}{2}\sum_{j=1}^J(\mu_j^2+\sigma_j^2-\ln(\sigma_j^2)-1).
\end{equation}
By updating the NN parameters to jointly reduce the cross entropy and the KL divergence $\mathbf{KL}[q(\mathbf{z}|\mathbf{y})||p(\mathbf{z})]$, we can both approximate the posterior and estimate the desired regression output.        

\section{Localization: View-Selective Deep Learning}
\label{sec:Localization: View-Selective Deep Learning}

\begin{figure}[h]
\begin{subfigure}{.32\textwidth}
    \centering
    \centerline{\includegraphics[width=4cm]{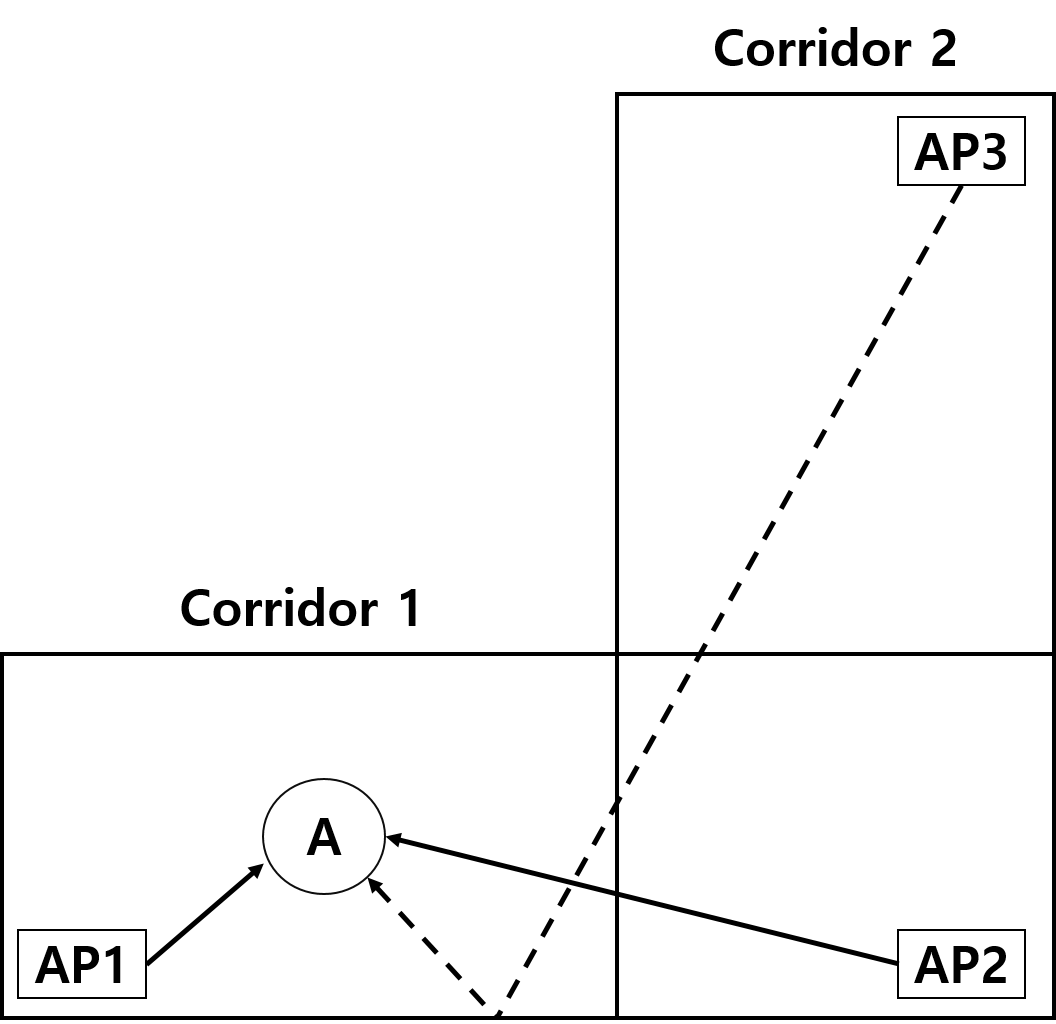}}
    \caption[]{}
%    \caption[Comparison with analysis (CDF)]{Comparison with analysis (CDF)
     \label{fig3}
\end{subfigure}
\begin{subfigure}{.32\textwidth} 
    \centering
    \centerline{\includegraphics[width=4cm]{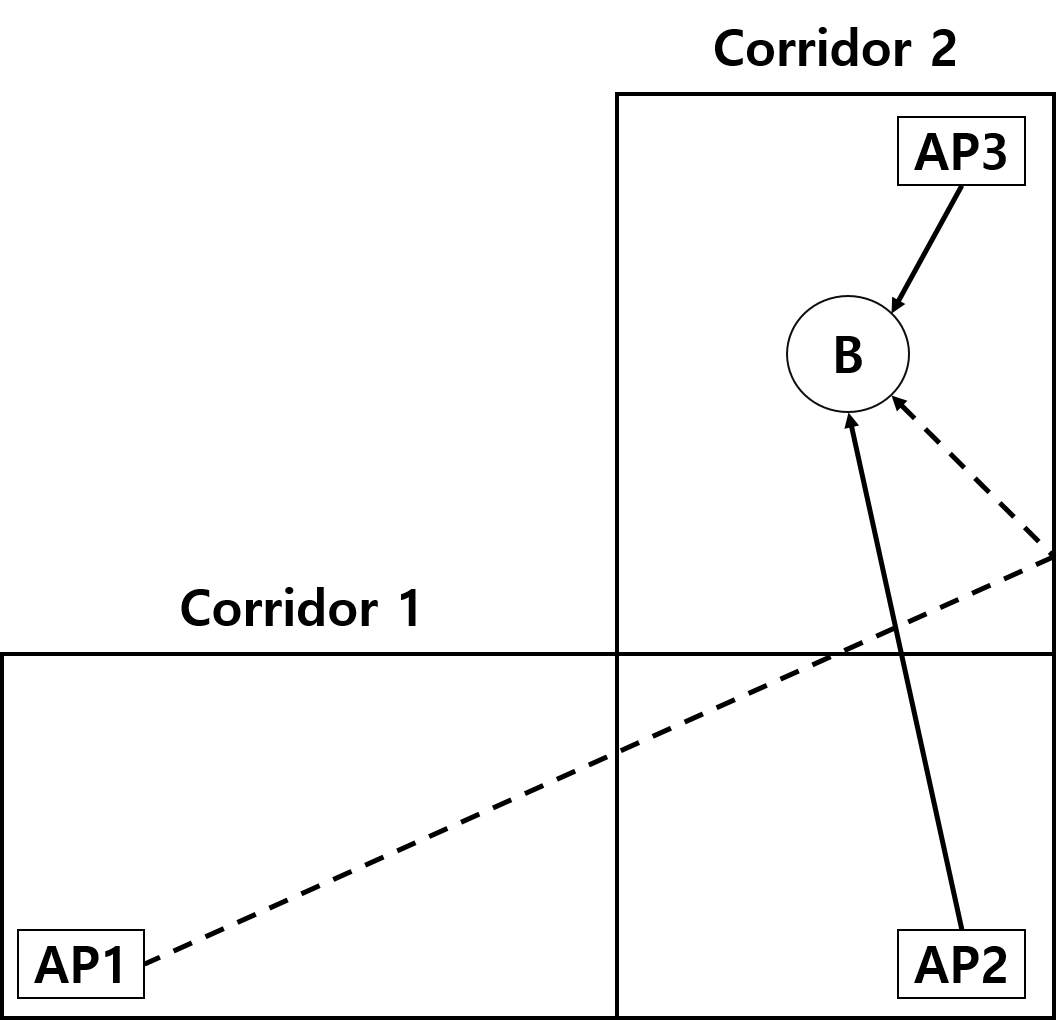}}
    \caption[]{}
%    \caption[Comparison with data type]{Comparison with analysis
     \label{fig4}
\end{subfigure}
\begin{subfigure}{.32\textwidth} 
    \centering
    \centerline{\includegraphics[width=4cm]{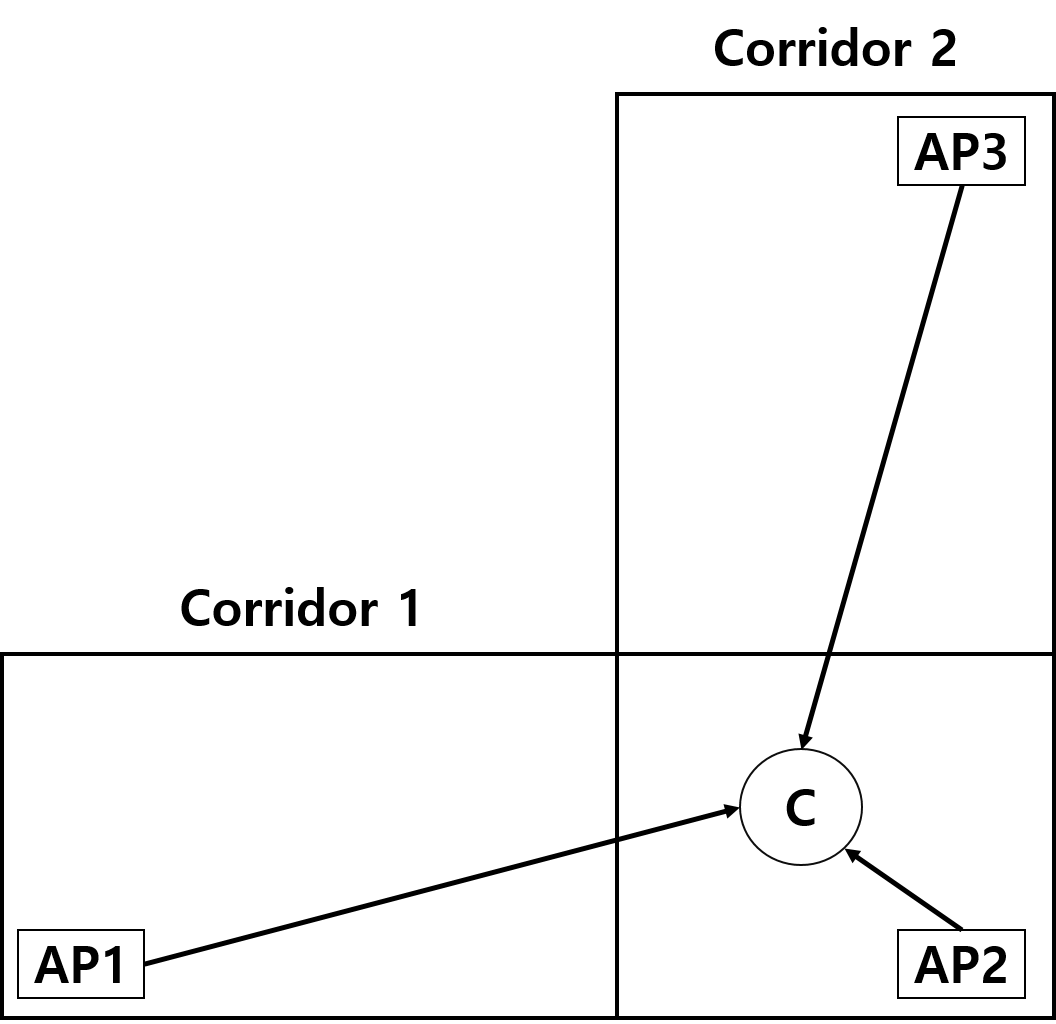}}
    \caption[]{}
%    \caption[Comparison with data type]{Comparison with analysis
     \label{fig5}
\end{subfigure}
\caption[Indoor localization on a two-corridor environment]{Indoor localization on a two-corridor environment. The radio signals from AP3 to A and AP1 to B are not on LoS. In terms of data views of APs, these NLoS signals make the corresponding views non-informative.}
\label{fig345}
\end{figure}

We introduce novel learning system, or the VSDL, with which the relative degree of importance of each view among the multiple views is used to improve the regression accuracy. A type of selective sampling, referred to as co-testing, was introduced \cite{muslea2000selective} to efficiently extract features from multiview data, where its basic idea is to inject divided views to multiple independent learning networks. However, this model had a strict assumption that each data sample might have a strong correlation among the views. On the other hand, we focus on a situation where views are not correlated to each other nor informative, but there is a given information whether or not a view is informative. In our case, we require localizing a target in a two-corridor environment, as in Figure \ref{fig345}. 

As you can see the Figure \ref{fig345}, the WiFi radio signal can propagate not only through a line of sight (LoS) but also non-line of sight (NLoS) paths. Subsequently the received signal consists of multi-path fading as well as the signal noise. The CSI data view at APs only from  NLoS paths are non-informative, since the multi-paths are very unpredictable. In this kind of case, it must be inefficient if all views are included in training. However, we can judge whether or not a view is informative so as to utilize such given information for supervised training. In our VSDL model, the learning parameters are much effectively updated by excluding non-informative data views. Looking into the big picture, VSDL is designed as a two-stage learning network for regression, consisting of a view-oriented variational deep network and a view-classified regression network. 

\subsection{View-Oriented Variational Deep Network}
\label{subsec:View-Oriented Variational Deep Network}

\begin{figure}
    \centering
    \centerline{\includegraphics[width=14cm]{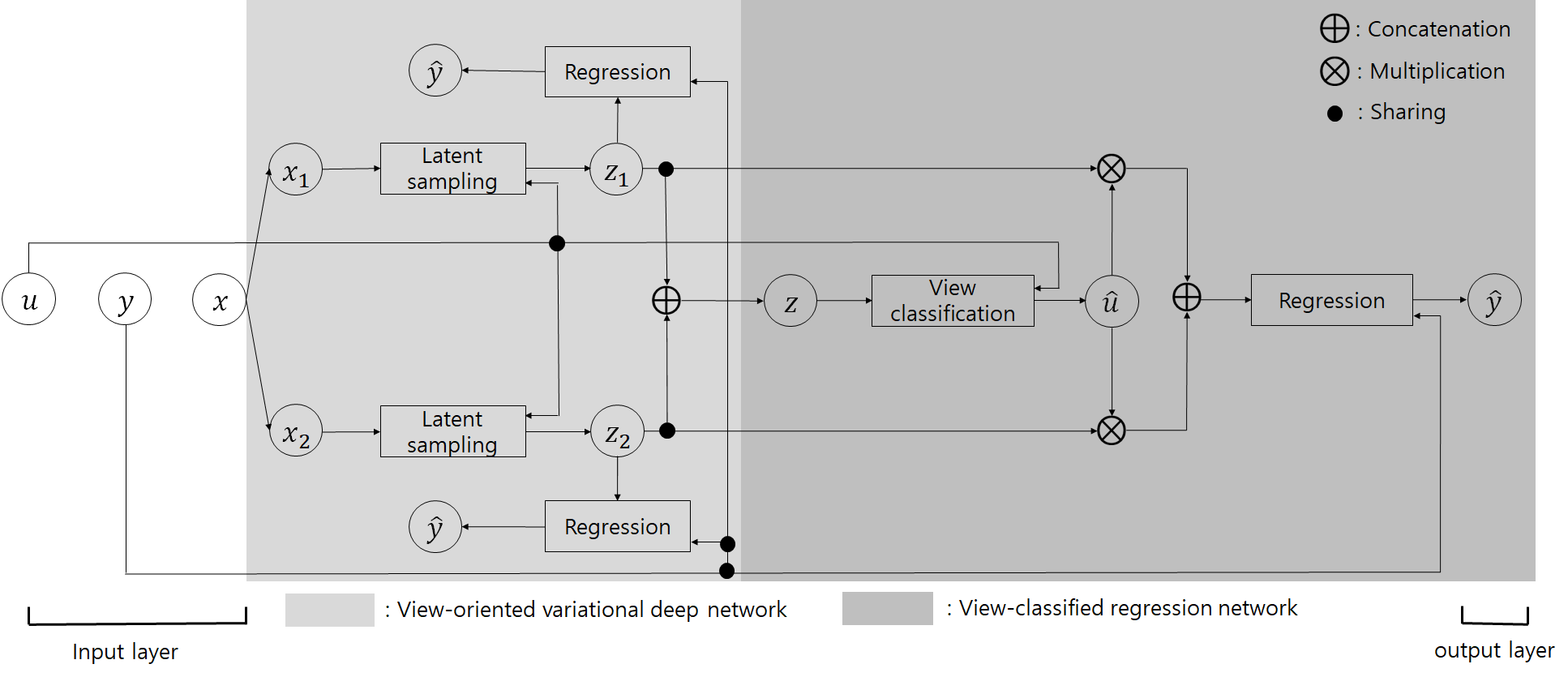}}
    \caption[]{Two-stage VSDL system design illustrated with an example of a two-view case. The view-oriented variational deep network (left grey box) consists of NNs for latent sampling and regression. The view-classified regression network (right grey box) consists of NNs for view classification and desired regression.}
%    \caption[Comparison with data type]{Comparison with analysis
     \label{fig12}
\end{figure}

The VSDL model is first trained to extract latent feature vectors from the multiview data in a view-oriented way, as shown in the left grey box in Figure \ref{fig12}. We first reconstruct the multiview input data $\mathbf{x}'=\{\mathbf{x}_1,\dots,\mathbf{x}_K\}$, from a single sample $\mathbf{x}$ to represent $K$ views. In our application, multiple views are created by different groups of WiFi APs located in different corridors. View $\mathbf{x}_k$ is the $k$-th subset of correlated data in $\mathbf{x}$. In our case, $K$ becomes the number of corridors and $\mathbf{x}_k$ consists of the data from the APs in the corridor $k$. Then we define $\mathbf{u}=\{u_1,\dots,u_K\}$, $u_k \in \{0,1\}$, that indicates the given view label over $K$ views. In our supervised training, we set the given corridor label $u_k=1$ if the location of the training sample is seen by a corridor view of $k$. Otherwise, set $u_k=0$. We model $K$ independent NNs to optimize the parameters. In short, each NN for the $k$-th view has a training input $(\mathbf{x}_k,\mathbf{y},u_k)$ as a set of data, true label, and given view label, respectively. In our case, usually $\mathbf{u}$ has a value of 1 for only one $k$, unless the training sample is seen by multiple views as seen in Figure \ref{fig5}, where it has a value of 1 for multiple $k$s. 

To apply variational inference, each latent sampling NN encodes $\mathbf{x}_k$ to latent feature $\mathbf{z}_k$ by the mapping, $\mathbf{z}_k = h_k(\mathbf{x}_k)$, with optimized mean-field vectors $\boldsymbol{\mu}_k=[\mu_{1,k},\dots,\mu_{J,k} ]$ and $ \boldsymbol{\sigma}_k  =[\sigma_{1,k},\dots,\sigma_{J,k} ]$ through $L$ hidden layers, $h^{(1)}_k, \dots, h^{(L)}_k$. The weight and bias $\mathbf{W}^{(l)}_k$ and $\mathbf{b}^{(l)}_k$ of the layer $l \in \{1,\dots,L\}$ are used to evaluate the feature output $\boldsymbol{\phi}^{(l)}_k$ ($\boldsymbol{\phi}^{(0)}_k = \mathbf{x}_k$) for the next layer input. The output layer $h^{(L)}_k$ generates the mean-field vectors: 
\begin{equation}\label{eq:8}
[\boldsymbol{\mu}_k, \boldsymbol{\sigma}_k]  = \mathbf{W}^{(L)}_k \boldsymbol{\phi}^{(L-1)}_k + \mathbf{b}^{(L)}_k,
\end{equation}
followed by reparameterization for latent vector $\mathbf{z}_k$ as
\begin{equation}\label{eq:9}
\mathbf{z}_k = \boldsymbol{\mu}_k + \boldsymbol{\sigma}_k \odot \boldsymbol{\epsilon}_k, ~~~\boldsymbol{\epsilon}_k \sim \mathcal{N}(0,\mathbf{I}).
\end{equation}
Here, we try to minimize KL divergence of $\mathbf{z}_k$ according to (\ref{eq:7}) to approximate the posterior $q(\mathbf{z}_k|\mathbf{y})$ to the distribution $p(\mathbf{z}_k)$.

Along with the process of the KL divergence minimization, the regression NN maps latent vector $\mathbf{z}_k$ to $\mathbf{\hat{y}}$ by the mapping, $\mathbf{\hat{y}} = g_k(\mathbf{z}_k|\mathbf{y})$, which consists of $P$ layers, $g^{(1)}_k, \dots, g^{(P)}_k$. The weight $\mathbf{W}'^{(P)}_k$ and bias $\mathbf{b}'^{(P)}_k$ of the last layer estimate the output $\mathbf{\hat{y}}= [\hat{y}_1,\hat{y}_2]$, which is represented in the normalized Cartesian coordinate in our case:
\begin{equation}\label{eq:10}
\mathbf{\hat{y}}  = \boldsymbol{\phi}'^{(P)}_k = \mathbf{W}'^{(P)}_k \boldsymbol{\phi}'^{(P-1)}_k + \mathbf{b}'^{(P)}_k.
\end{equation}
Then its regression loss is evaluated in terms of Euclidean distance with the known true $y$ for supervised learning. We jointly minimize the KL divergence and the regression loss by updating the weight and bias parameters. 
 
In order to achieve supervised view-oriented learning, we utilize the given view label $\mathbf{u}$ during the training. For every multiview training sample, the mappings $h_k(\mathbf{x}_k)$ and $g_k(\mathbf{z}_k|\mathbf{y})$ are optimized to generate $\mathbf{z}_k$ only for the informative views as follows:
\begin{equation}\label{eq:11}
\left\{ \begin{array}{ll}
         %\mbox{Jointly minimize loss and update  $\mathbf{W}^{(l)}_k$, $\mathbf{b}^{(l)}_k$, $\mathbf{W}'^{(p)}_k$, and $\mathbf{b}'^{(p)}_k$} & \mbox{if $u_k == 1$};\\
            \min_{\mathbf{W}_k,\mathbf{b}_k,\mathbf{W}'_k,\mathbf{b'}_k}
            {\{(y_1-\hat{y}_1)^2+(y_2-\hat{y}_2)^2\}+\mathbf{KL}[q(\mathbf{z}_k|\mathbf{y})||p(\mathbf{z}_k)]} & \mbox{if $u_k = 1$},  \\
         \mbox{Do nothing} & \mbox{if $u_k = 0$}.
\end{array} \right.
\end{equation}
We expect the NNs properly extract latent features with excluding the non-informative data view. Although trained weights and biases can generate the latent feature vector $\mathbf{z}_k$ for every test sample regardless of the condition $u_k$, the invisible latent features from two different samples may have a strong correlation if both are informative in a certain view $k$ (i.e., $u_k=1$). Now $\mathbf{z}$ for every training and test sample becomes a new input in the next step. In the following section \ref{subsec:View-Classified Regression Network}, we will introduce an additional network to enhance regression using the latent features and their invisible correlation.  

\subsection{View-Classified Regression Network}
\label{subsec:View-Classified Regression Network}

The view-classified regression network, as described in the right grey box in Figure \ref{fig12}, uses an intermediately integrated \cite{noble2004support} latent vector $\mathbf{z} = \{\mathbf{z}_1,\dots,\mathbf{z}_K\}$ and the given view label $\mathbf{u}$ used in section \ref{subsec:View-Oriented Variational Deep Network}. The network consists of two NNs; 1) to classify view information $\mathbf{\hat{u}}$ and 2) to obtain regression output $\mathbf{\hat{y}}$ regarding each classified view information $\hat{u}_k$ as a reweight parameter to a subset $\mathbf{z}_k$. Our insight in this network starts from the hypothesis that the latent vector $\mathbf{z}$ generated from the previous view-oriented learning can select dominant views $k$s through the view classification NN. The regression NN then utilizes the classification result to enhance the desired regression performance. The aim is to jointly approximate the classification output $\mathbf{\hat{u}}$ to the given view label $\mathbf{u}$, and the regression output $\mathbf{\hat{y}}$ to the true label $\mathbf{y}$. 
The classification output $\mathbf{\hat{u}} = [\hat{u}_1,\dots,\hat{u}_K]$ becomes the reweight parameter where $\hat{u}_k$ means how importantly the regression NN should consider the influence of view $k$. First, since more than one $u_k$s may have a value of 1, we must regulate them with normalization to feed view classification NN with to learn the balanced reweight parameter:  
\begin{equation}\label{eq:12}
\tilde{u}_k = \frac{u_k}{\Sigma_{i=1}^K u_i}.
\end{equation}
The view classification NN of layers $h^{(1)}_Q,\dots,h^{(Q)}_Q$ maps $\mathbf{z}$ to $\mathbf{\hat{u}}$ such that $\mathbf{\hat{u}} = h_Q(\mathbf{z}|\mathbf{\tilde{u}})$. Starting from the first layer input $\mathbf{z}$ ($\boldsymbol{\phi}^{(0)}_Q = \mathbf{z}$), we calculate the view classification result $\mathbf{\hat{u}}$ from the output layer through the softmax activation:  
\begin{equation}\label{eq:13}
\mathbf{\hat{u}} =  \boldsymbol{\phi}^{(Q)}_Q = 
\mbox{softmax}(\mathbf{W}^{(Q)}_Q \boldsymbol{\phi}^{(Q-1)}_Q + \mathbf{b}^{(Q)}_Q),
\end{equation}
where $\mathbf{W}^{(q)}_Q$ and $\mathbf{b}^{(q)}_Q$ denote the weight and bias of the layer $q \in \{1,\dots,Q\}$, respectively.

Regarding $\mathbf{\hat{u}}$ as the reweight parameter, we recalculate each subset $\mathbf{z}_k$ to $\mathbf{z'}_k$:
\begin{equation}\label{eq:14}
\mathbf{z'}_k = \hat{u}_k \odot \mathbf{z}_k,
\end{equation}
and use it as a regression input. With concatenated $\mathbf{z}'= \{\mathbf{z'}_1,\dots,\mathbf{z'}_K\}$, the regression NN of layers $h^{(1)}_R,\dots,h^{(R)}_R$ maps $\mathbf{z}'$ to $\mathbf{\hat{y}}$, such that $\mathbf{\hat{y}} = h_R(\mathbf{z}'|\mathbf{y})$, and obtains the Cartesian coordinate output $\mathbf{\hat{y}} = [\hat{y}_1,\hat{y}_2]$, in our case:
\begin{equation}\label{eq:15}
\mathbf{\hat{y}} = \boldsymbol{\phi}^{(R)}_R = \mathbf{W}^{(R)}_R \boldsymbol{\phi}^{(R-1)}_R + \mathbf{b}^{(R)}_R,
\end{equation}
where $\mathbf{W}^{(r)}_R$ and $\mathbf{b}^{(r)}_R$ are the weight and bias of the layer $r \in \{1,\dots,R\}$. The NNs update parameters $\mathbf{W}_Q$, $\mathbf{b}_Q$, $\mathbf{W}_R$, and  $\mathbf{b}_R$ to jointly minimize both Euclidean losses:  
\begin{align}\label{eq:16}
\min_{\mathbf{W}_Q,\mathbf{b}_Q,\mathbf{W}_R,\mathbf{b}_R}
\alpha\{(y_1-\hat{y}_1)^2+(y_2-\hat{y}_2)^2\}+(1-\alpha)\{\Sigma_k{(\tilde{u}_k-\hat{u}_k)^2}\},   
\end{align}
where $\alpha \in (0,1)$ denotes a trade-off parameter between two losses. With the fingerprint database consisting of weights and biases, for the test data, we can obtain the regression output $(\hat{y}_1,\hat{y}_2)$ which is the localization result. Strictly speaking, our reweighting is different from the existing iterative reweight (IR) methods \cite{chartrand2008iteratively,mohan2012iterative}, which sought the reweighting based on their gradient directions. In contrast, we suggest a simpler method that derives reweight parameters based on the given information. Our reweighting method assists the supervised learning system to decide which data view should be considered more importantly and further to achieve better performance.

\section{Field Experiment}

\begin{figure}[h]
\begin{subfigure}{.32\textwidth}
    \centering
    \centerline{\includegraphics[width=4.3cm]{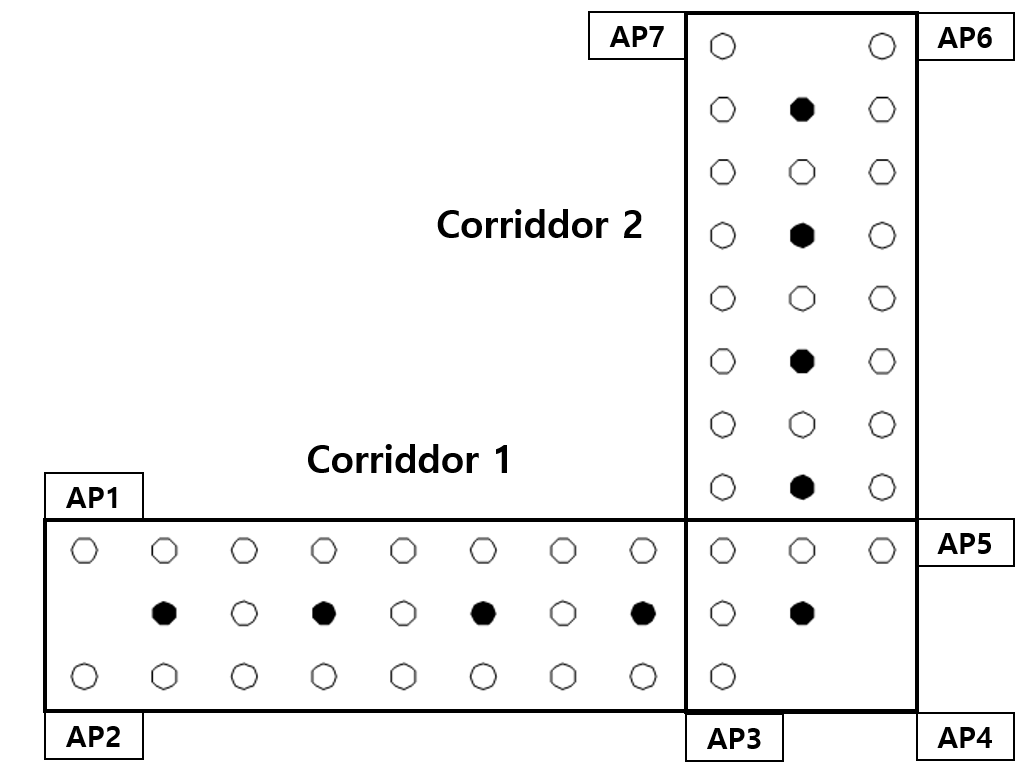}}
    \caption[]{}
%    \caption[Comparison with analysis (CDF)]{Comparison with analysis (CDF)
     \label{ex_fig0}
\end{subfigure}
\begin{subfigure}{.32\textwidth}
    \centering
    \centerline{\includegraphics[width=4.3cm]{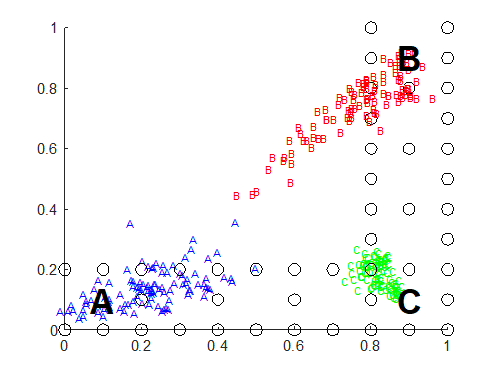}}
    \caption[]{}
%    \caption[Comparison with analysis (CDF)]{Comparison with analysis (CDF)
     \label{ex_fig2}
\end{subfigure}
\begin{subfigure}{.32\textwidth} 
    \centering
    \centerline{\includegraphics[width=4.3cm]{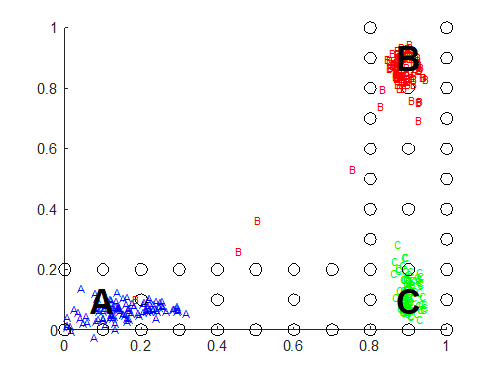}}
    \caption[]{}
%    \caption[Comparison with data type]{Comparison with analysis
     \label{ex_fig1}
\end{subfigure}

\caption[Two-corridor experiment environment]{Two-corridor experiment. (a) Localization topology with 43 training points (open circle) and 9 test points (solid circle). There are seven APs to receive packets from the points at the same time. (b) Regression results from a variational DL system with no exclusion of non-informative views. (c) Regression results from the VSDL system with exclusion of non-informative views, which finds the dominant views effectively to enhance localization accuracy.
\vspace{-1em}
}
\label{Two-corridor experiment environment}
\end{figure}

We apply the VSDL system for indoor localization on a two-corridor real building environment with 43 training points and 9 test points, as in Figure \ref{ex_fig0}. Each corridor is 7 m long, where training and test points are spread by 0.5 m spacing. We install seven APs consisting of 3-antenna IWL 5300 NIC in laptop computers, which are placed at the corners of the corridors. For a transmitter, the same laptop with a single antenna is used to transmit WiFi packets on channel 36 at 5.18 GHz. The APs receive packets from the transmitter at the same time using the monitor mode and we combine them as a multiview input at the server side. Each AP receives the WiFi packet with three antennas to form three Tx-Rx radio channels. Each channel produces a CSI vector consisting of 30 subcarrier CSIs ($I=30$). We then take one of the three CSI vectors as the reference to produce two relative CSI vectors. In this way, we obtain an input sample $\mathbf{x}$ consisting of 420 $(=7\times(3-1)\times30)$ relative CSIs. We collect 100 sample packets for every training and test points in a noisy environment, which manifests high data fluctuation that makes it difficult to estimate the location by other analytical methods.

In this scenario, we divide input $\mathbf{x}$ into two views $\mathbf{x_1}$ and $\mathbf{x_2}$ ($K=2$), which represent AP associations with corridors 1 and 2, respectively. Therefore, $\mathbf{x_1}$ has information of AP1 to AP5 and $\mathbf{x_2}$ has that of AP3 to AP7. Along with the CSI data, the location label $\mathbf{y}$ in the Cartesian coordinate and the given corridor label $\mathbf{u}$ are utilized for training. There are three cases for $\mathbf{u}$ depending on the training location: The location belongs, 1) only to corridor 1 ($\mathbf{u}=[1,0]$), 2) only to corridor 2 ($\mathbf{u}=[0,1]$), and 3) to both corridors 1 and 2 ($\mathbf{u}=[1,1]$). The NNs for view-oriented learning update parameters only when the view is informative ($u_k=1$). Here, the information from AP3 to AP5 are common in both views and considered to be informative for every sample.

\begin{figure}[h]
\centering
\begin{minipage}{.48\textwidth}
    \centering
    \centerline{\includegraphics[width=5cm]{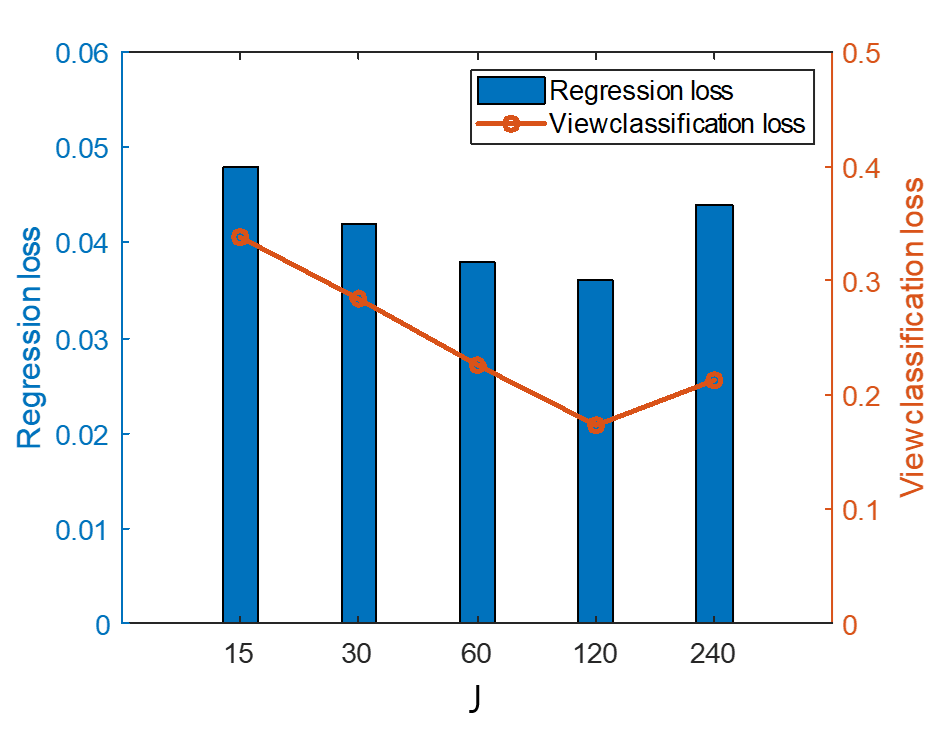}}
    \captionof{figure}{Regression and view classification losses versus $J$.  }\label{ex_fig4}
\end{minipage}\hfill\hfill\hfill\hfill
\begin{minipage}{.48\textwidth}
    \centering
    \centerline{\includegraphics[width=5cm]{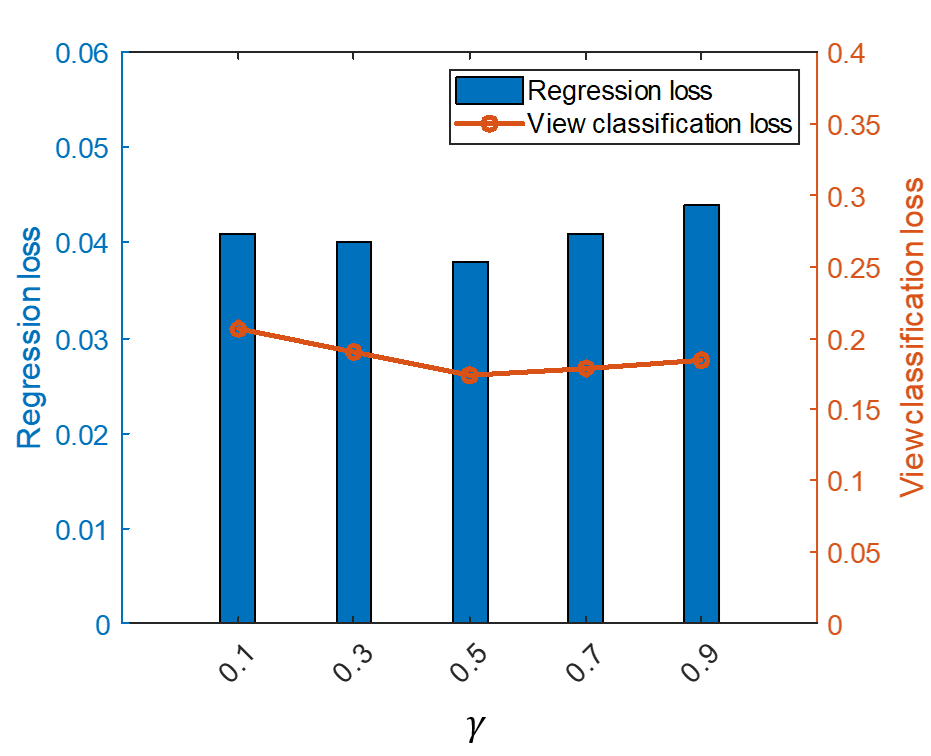}}
    \captionof{figure}{Regression and view classification losses versus $\alpha$.} \label{ex_fig5}
\end{minipage}
%\setlength{\belowcaptionskip}{-200pt}
%\vspace{-1em}
\end{figure}

Along the entire system, we set the number of hidden layers $L$, $P$, $Q$, and $R$ to be three. The number of nodes of each layer decreases from 1000 to 500. For the feature output of every layer except for the last one, the Relu activation is used. Adam optimizers are used to update the parameters with learning rate of $10^{-5}$. First, we aim to verify if our joint optimization works well. As seen in Figure \ref{ex_fig4}, the losses of regression and view classification are minimized together for any given number of latent variable $J$. This trend implies that our system properly operates to obtain an advanced regression result assisted by the view selection. In addition, we obtain the best regression result with $J=120$ rather than keeping the size of the input CSI vector, which corresponds to a mean-field feature compression ratio of $2/7$. There should be a sufficiently large number of variables to estimate the posterior of the CSI data, while too many variables cause overfitting of the network. Obviously, $J$ depends on the scenario as well as the application requirement.

In our VSDL system, the trade-off parameter $\alpha$ can influence the performance of regression as seen in Figure \ref{ex_fig5}. As $\alpha$ approaches 0, the network becomes too sensitive on view classification, and it makes both losses worse. On the other hand, as $\alpha$ approaches 1, high view classification loss occurs, resulting in poor regression (localization) accuracy. We obtain the best regression accuracy with $\alpha=0.5$.        

Figures \ref{ex_fig2} and \ref{ex_fig1} show comparisons of the regression results from a simple variational DL and our VSDL with $J=120$ and $\alpha=0.5$. We take three test locations as the representative cases. The locations A and B are near the end of corridors and location C is in the intersection of the two corridors. The test results for the location A, B, and C are shown in blue, red, and green, respectively. In terms of multiview data learning, the variational DL extracts features from all corridor views including non-informative views. Therefore, as seen in Figure \ref{ex_fig2}, the regression results for many cases are located somehow outside the topology of training. In contrast, the VSDL system updates the learning parameters only for informative views to classify the dominant view and hence to achieve better localization, as seen in Figure \ref{ex_fig1}.

\begin{minipage}[h]{0.48\textwidth}
    \centering
    \centerline{\includegraphics[width=6cm]{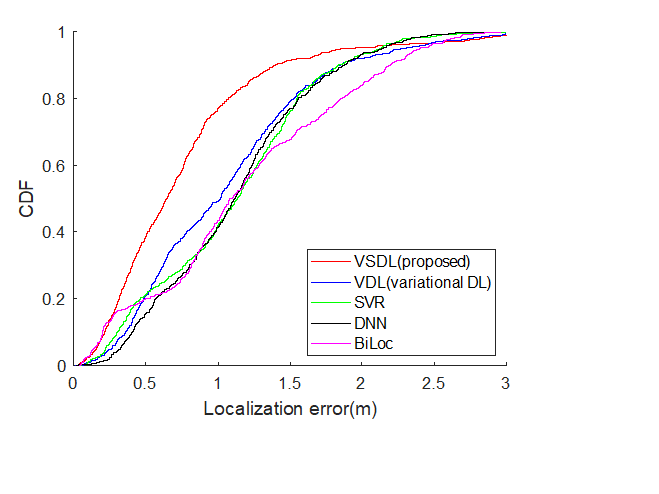}}
%    \rule{6.4cm}{3.6cm}
    \vspace{-2em}
    \captionof{figure}{Localization error CDF of the systems. The proposed VSDL significantly outperforms other previous systems. }\label{CDF}
   
\end{minipage}
\hfill
\begin{minipage}[h]{0.48\textwidth}
    \centering
    \vspace{3.7em}
    {\small
    \begin{tabular}{cc}\hline
      Algorithm & Localization error (m) \\ \hline
        VSDL(proposed)  & 0.7715 \\
        VDL (variational DL) & 1.0607 \\
        SVR & 1.1037 \\
        DNN & 1.1246 \\
        BiLoc & 1.1844 \\
        CiFi & 1.9739 \\ \hline
    \end{tabular}
    \captionof{table}{Localization error comparison. The proposed VSDL improves 30 $\%$ of the localization accuracy}\label{table}  
    }
\end{minipage}
%  \end{minipage}

Further, we compare the proposed VSDL with several existing machine learning systems. To discriminate the CSI data, both classification and regression methods were introduced in previous papers to improve the localization accuracy. In our experiment environment, as well as simple variational DL, we implemented RBM based classification BiLoc \cite{wang2017biloc}, CNN based classification CiFi \cite{wang2017cifi}, SVM based regression SVR \cite{zhou2017csi}, and stand-alone DNN based regression. Figure \ref{CDF} and Table \ref{table} show the comparison results. We do not plot the results of CiFi in this scenario, since the convolutional analysis from batch information cannot extract proper features and results in a very poor localization accuracy of 1.97 m. First, the variational DL, whose results are described in \ref{ex_fig2}, outperforms other existing systems due to the usage of variational inference. Here, we observe that the introduction of the variational inference brings the key advantage for WiFi CSI localization in a noisy radio channel. In addition, the VSDL system with novel two-stage view-selective learning on the variational inference base significantly improves the localization accuracy by 30 $\%$, from 1.10 m to 0.77 m. As the VSDL is very scalable by the nature of its design, we expect further performance improvement in environments with more corridors. 

\vspace{-0.5em}
\section{Conclusions}
\vspace{-0.5em}

WiFi device localization has been a very attractive area of study as WiFi networks are omnipresent to provide network application services to anonymous users in these days, anticipated to open new business opportunities as well as new technical challenges. The technical performance of WiFi localization has been improved disruptively by the use of channel state information measured at multiple receive APs. 

In this paper, we introduce a machine learning design that combines the variational deep learning very effectively in multiview learning architecture. We report an observation that the latent vectors generated at the intermediate layer of variational deep learning form strong feature behaviors to provide classification of effective view selection that enhance the accuracy of localization by a great deal. Our system, the view-selective deep learning, or the VSDL, achieves a localization accuracy of 0.77 m, which manifests a more than 30 \% improvement in a two-corridor field experiment compared with the best known system based on SVM. The VSDL is completely scalable as exploiting the benefit of multiview-based regression, and hence the WiFi localization network can be expanded with no limit, such as in complex building structure. Our design of extracting features in the latent space to deal with informative and non-informative views in a multiview variational deep learning network is very powerful so that it can be applied to various applications with no limit on scalability. 

%\newpage
\section*{Broader Impact}

The indoor localization with radio signals, for example, WiFi radio signals, which finds the location of a mobile device very accurately can create a great deal of impact in mobile service application. This can be a benefit to off-line stores and services such as in a shopping mall, hospitial, and public buildings, where location-based services can directly improve quality of experiences, especially when associated with social network services. Of course, such network features of localization can also harm the privacy of people in public. Radio localization may fail in a hot spot in the sense of crowded radio network traffic. However, such a failure may not cause critical problems except for some frustration with internet-based applications. 

%Authors are required to include a statement of the broader impact of their work, including its ethical aspects and future societal consequences. 
%Authors should discuss both positive and negative outcomes, if any. For instance, authors should discuss a) who may benefit from this research, b) who may be put at disadvantage from this research, c) what are the consequences of failure of the system, and d) whether the task/method leverages biases in the data. If authors believe this is not applicable to them, authors can simply state this.

%Use unnumbered first level headings for this section, which should go at the end of the paper. {\bf Note that this section does not count towards the eight pages of content that are allowed.}

%References follow the acknowledgments. Use unnumbered first-level heading for
%the references. Any choice of citation style is acceptable as long as you are
%consistent. It is permissible to reduce the font size to \verb+small+ (9 point)
%when listing the references.
%{\bf Note that the Reference section does not count towards the eight pages of content that are allowed.}
%\medskip
%%
\small
\bibliographystyle{unsrt}
\bibliography{ref}

%[1] Alexander, J.A.\ \& Mozer, M.C.\ (1995) Template-based algorithms for
%connectionist rule extraction. In G.\ Tesauro, D.S.\ Touretzky and T.K.\ Leen
%(eds.), {\it Advances in Neural Information Processing Systems 7},
%pp.\ 609--616. Cambridge, MA: MIT Press.
%
%[2] Bower, J.M.\ \& Beeman, D.\ (1995) {\it The Book of GENESIS: Exploring
%  Realistic Neural Models with the GEneral NEural SImulation System.}  New York:
%TELOS/Springer--Verlag.
%
%[3] Hasselmo, M.E., Schnell, E.\ \& Barkai, E.\ (1995) Dynamics of learning and
%recall at excitatory recurrent synapses and cholinergic modulation in rat
%hippocampal region CA3. {\it Journal of Neuroscience} {\bf 15}(7):5249-5262.

\end{document}